# Building AI-Ready Data Systems for Space Life Sciences, Aerospace Medicine, and Deep Space Exploration

Sylvain V. Costes[1,2,3*], Sergio Garcia Busto[4,5,6], Ryan T. Scott[7], James A. Casaletto[1], Gautier Bardi de Fourtou[8], Brian M. Evarts[9] , Amanda M. Saravia-Butler[7], Xavier-Lewis Palmer[10], Rodrigo Coutinho de Almeida[11], Laetitia Frost[11], Jelena Tešić[12], Afshin Beheshti[2,3,13,14,15,16], Christopher E. Mason[3,17], Peter W. Rose[18], Sergio E. Baranzini[19], Lauren M. Sanders[1], Stefania Giacomello[20], Pedro Madrigal[21*]

[1] Blue Marble Space, Seattle, Washington, USA
[2] Computational and Systems Biology, University of Pittsburgh, Pittsburgh, PA, USA
[3] Trivedi Institute for Space and Global Biomedicine, University of Pittsburgh, Pittsburgh, PA, USA
[4] Wellcome Sanger Institute, Wellcome Genome Campus, Hinxton, CB10 1SA, UK
[5] University of Cambridge, Cambridge, UK
[6] Department of Systems Biology, Harvard Medical School, Boston, MA, USA
[7] Amentum, Space Biosciences Division, NASA Ames Research Center, Moffett Field, CA, USA
[8] Massachusetts Institute of Technology, Cambridge, USA
[9] Crown Point Technologies, Inc, Columbia, MD, 21046
[10] BiosView Labs, Dayton, Ohio, USA
[11] Directorate of Human and Robotic Exploration, European Space Agency, Noordwijk, the Netherlands
[12] Texas State University, USA
[13] McGowan Institute for Regenerative Medicine, Department of Surgery, University of Pittsburgh School of Medicine, Pittsburgh, PA, USA
[14] Center for Space Biomedicine, Trivedi Institute for Space and Global Biomedicine, University of Pittsburgh School of Medicine, Pittsburgh, PA, USA
[15] Department of Bioengineering, University of Pittsburgh, Pittsburgh, PA, USA
[16] Stanley Center for Psychiatric Research, Broad Institute of MIT and Harvard, Cambridge, MA, USA
[17] Department of Systems and Computational Biomedicine and WorldQuant Initiative, Weill Cornell Medicine, New York, NY, USA
[18] San Diego Supercomputer Center, University of California San Diego, La Jolla, CA, USA
[19] Weill Institute for Neurosciences, Department of Neurology, University of California San Francisco, USA
[20] Science for Life Laboratory, Department of Gene Technology, KTH Royal Institute of Technology, Stockholm, Sweden
[21] European Molecular Biology Laboratory, European Bioinformatics Institute (EMBL-EBI), Hinxton CB10 1SD, UK

* To whom correspondence should be addressed: pmadrigal@ebi.ac.uk, syc63@pitt.edu

## Abstract

While AI holds the potential to revolutionize space life sciences, realizing this promise is contingent upon the systematic restructuring of heterogeneous spaceflight biological data into machine-actionable, AI-ready forms. Even though open access principles support human reuse and scientific reproducibility, this does not necessarily enable AI systems to access and analyze such a diverse set of scientific datasets. In addition, the growing array of AI approaches places distinct demands on data structure, metadata, and access interfaces. In order to respond to such growing changes we propose a three-tier approach, proceeding from FAIR to AI-ready to space-ready data. We discuss existing infrastructures and how they can be improved to close the AI access gap. We conclude by proposing a neutral international coordinating body as the governance backbone for the trustworthy, agent-accessible space biology infrastructure that deep space biological research will require.

## Introduction

Space life sciences study how living organisms respond to spaceflight conditions, generating the foundational knowledge for the applied science and operational solutions required for long-duration missions[1]. The sophistication of biological experimentation has advanced incrementally with the introduction of new hardware, including automated cell culture systems, miniature bioreactors, sequencing technologies, and advanced microscopy. The adoption of these tools has been critical in discovering the effects of the unique space environment on organismal biology and physiology, including oxidative stress, DNA damage, mitochondrial dysregulation, epigenetic changes, telomere length alterations, and microbiome shifts, but their precise interplay under space conditions remains to be elucidated[1]. Spaceflight research now spans diverse fields such as cancer, aging, neuroscience, immunology, and pharmacology, and the data generated reach well beyond omics: phenotypic measurements, imaging, radiation dosimetry, and digital biomarkers from wearable devices are now routinely collected alongside molecular profiles.
The growing volume of spaceflight data, facilitated by the expansion of space omics, led to the creation of GeneLab and ultimately NASA's Open Science Data Repository (OSDR), a unified infrastructure that aggregates heterogeneous biological, hardware, and environmental data[2,3]. OSDR contains Life Sciences data from Ames Life Sciences Data Archive (physiological, phenotypic, imaging, environmental telemetry data), GeneLab (Molecular/Omics Data) and NBISC (Biospecimens).

The NASA Artemis program, which plans sustained lunar presence and crewed transit to Mars[4], together with an expanding commercial spaceflight sector, has the potential to generate biological and environmental datasets at a scale and cadence the field has not previously seen[5,6]. Realizing that potential requires transitioning from isolated experiments toward integrated analyses that can support autonomous biological research and life-support operations/systems under microgravity and ionizing radiation[7–9]. NASA has formally identified this need, defining critical technology gaps that include a crew health and performance integrated data architecture, autonomous monitoring for exploration, and trustworthy autonomy for planning and decision-making, each likely requiring AI capable of reasoning over heterogeneous life science data[10].

AI performs well at learning patterns in biological data, yet adoption across space life sciences remains nascent. The barriers include extreme data scarcity (low subject *n*), high heterogeneity across modalities and model organisms (samples span from Drosophila to humans), data quality (e.g. noisy datasets driven by spaceflight resource constraints[11]), paucity of sex-balanced studies[9,12], and fragmented data access[12].

However, even with limited adoption, AI implementation has shown great promise for space life sciences research. In Table 1, we categorize AI methodologies and show their utility in addressing specific spaceflight constraints. Briefly, generative approaches (e.g., foundation models and agentic AI) coupled with retrieval augmented generation (RAG) could facilitate transfer learning from the broad Earth-based biological knowledge base for small-*n* spaceflight prediction; network-based methods (knowledge graphs, graph machine learning) can map complex biological interactions within the unique spaceflight environment (e.g., elevated CO2, radiation, psychological factors); and architectural frameworks such as federated learning would preserve astronaut data privacy[7–9]. Classical machine learning (classification and regression) and well-characterized computer vision models are the building blocks of an AI-based biological discovery toolkit, including causal inference frameworks to move beyond correlation and disentangle the distinct effects of microgravity and radiation despite limited sample sizes[28]; and finally, reinforcement learning would enable efficient, low-power inference for limited in situ computing on space hardware; as well as provide appropriate training for simulation models and digital twins for personalized responses to mission stressors[7–9]. Table 1 summarizes how these methods impose distinct demands on data structure, metadata, and access interfaces[13]. The research approach is co-defined by the scientific objective and the nature of the data; this intersection determines the appropriate AI methodology and the subsequent standards for data readiness. A federated learning model trained on cross-cohort spaceflight omics faces different requirements than a foundation model (FM) fine-tuned on spaceflight transcriptomes, and both differ from a vision model classifying microscopy or optical coherence tomography data.

| **AI method category** | **Representative spaceflight application** | **Primary data requirement (space biology context)** | **AI-readiness Component Examples** | **Exemplar reference** |
|---|---|---|---|---|
| **Foundation models (FMs) and Large Language Models (LLM)** | Fine-tuning biomedical FMs (e.g., Geneformer and scGPT for single-cell transcriptomics, TabPFN for tabular spaceflight data) for spaceflight tasks; pre-training on space biology literature and multi-omics corpora; transfer from terrestrial domains for small-*n* spaceflight prediction tasks. | Sufficient corpus scale against terrestrial biomedicine; harmonized multimodal tokenization; auditable separation of pretraining and evaluation data; downstream task labels for fine-tuning. | Cross-study harmonization and rich metadata (AI-Readiness Requirements); domain transfer pipelines and e.g. Hugging Face Hub distribution (Access Layer). | Theodoris et al, 2023[14] *biomedical*<br>Cui et al, 2024[15] *biomedical*<br>Hollmann et al, 2025[16] *ML* |
| **Agentic AI** | Agentic orchestration across OSDR, Monarch, PubMed, and RadLab to compose multi-step analyses autonomously. | Programmatic interfaces per archive; machine-queryable schemas; auditable provenance for agent reasoning chains. | MCP servers and agent-accessible query layers (*Access Layer*); provenance enforcement (*AI-Readiness Requirements*). | Li et al, 2026[17] *biomedical* |
| **Retrieval-augmented generation (RAG)** | Grounding LLM outputs in OSDR metadata and study methods via retrieval over semantically chunked indices; KG-augmented RAG across spaceflight omics; LLM-assisted biocuration over LinkML schemas. | Semantically chunked metadata; LLM-indexable schemas; indexes kept current with curation events. | Structured metadata and LinkML schemas (*AI-Readiness Requirements*); MCP-mediated retrieval over SPOKE-GeneLab (*Access Layer*). | Soman et al, 2024[18] *biomedical*<br>Caufield et al, 2024[19] *biomedical* |

| AI method category | Representative spaceflight application | Primary data requirement (space biology context) | AI-readiness Component Examples | Exemplar reference |
|---|---|---|---|---|
| **Knowledge graphs (KGs) and graph ML** | Integrating OSDR multi-omics, clinical, and environmental data via curated KGs such as the GeneLab KG (federated with SPOKE through the SPOKE-GeneLab proxy-node architecture) and PrimeKG; graph FMs for cross-mission hypothesis generation and drug repurposing; GNN inference over biomedical entity networks. | Entity and relation schemas; ontology integration across missions and domains; LinkML-aligned structured assertions in place of free-text metadata; persistent identifiers for nodes and edges. | LinkML data modelling and ontology integration (*AI-Readiness Requirements*); GeneLab KG with federated SPOKE-GeneLab access (*Access Layer*). | Nelson et al, 2021[20] *spaceflight*<br>Chandak et al, 2023[21] *biomedical*<br>Huang et al, 2024[22] *biomedical* |
| **Federated learning** | Training across distributed cohorts (NASA and commercial spaceflight, analog studies, terrestrial biobanks) without centralizing restricted-tier data; ground-to-space federation on orbital compute. | Harmonized features across sites; pre-federation quality gates; secure aggregation and differential privacy; cross-jurisdiction governance. | Federated architectures with AIDRIN pre-federation scoring and governance threading (*Federated Learning*). | Casaletto et al, 2025[23] *spaceflight*<br>Sheller et al, 2020[24] *biomedical*<br>Pati et al, 2022[25] *biomedical* |
| **Computer vision** | Pose estimation and behavioral segmentation of ISS-housed rodents; crew video analysis for ambulation, task performance, and team dynamics; classification of bacteria in microscopy samples; SANS retinal imaging analysis. | Annotated ground truth at scale; robustness to mission-induced imaging artifacts (occlusion, soiling, aberration); limited labeled data per mission. | Cross-study harmonization and transfer learning from terrestrial cohorts (*AI-Readiness Requirements*); medical-imaging FMs for zero-shot segmentation (*Access Layer*). | Pereira et al, 2022[26] *biomedical*<br>Bohnslav et al, 2021[27] *biomedical*<br>Ma et al, 2024[28] *Biomedical*<br>Huang et al, 2025[29] *spaceflight* |
| **Classical machine learning / Causal inference** | Feature discovery from low-sample-size, high-dimensional omics; tabular predictors on mission manifest and physiological data. | Batch-effect-aware harmonization across missions; $p \gg n$ regime; rich covariate metadata; principled handling of missingness and any imputation. | Batch-correction pipelines (*AI-Readiness Requirements*); metadata schemas enabling cross-study pooling (*AI-Readiness Requirements*). | Casaletto et al, 2025[30] *spaceflight*<br>Ilangovan et al, 2024[11] *spaceflight* |
| **Reinforcement learning** | Adaptive experimentation in self-driving laboratories; closed-loop control of in-flight biological experiments; sample-efficient experimental design under spaceflight resource constraints. | Simulation environments or high-fidelity surrogates; reward definition in operational contexts; offline learning from sparse observational data. | Digital twins and surrogate models (*Priorities and Outlook*); adaptive autonomy framing (*Access Layer*). | Gottesman et al, 2019[31] *biomedical* |

**Table 1. AI method categories and their data requirements in space life sciences.** Each of these eight AI approaches stresses different characteristics of the underlying data; no single dataset can satisfy all of them simultaneously. Rows pair each method with a representative spaceflight application, the principal data requirement in space biology, the AI-readiness lever addressed in the main text (italicized section name), and an exemplar whose domain tag (spaceflight, biomedical [terrestrial], or ML) indicates where published precedent currently exists. AIDRIN = AI Data Readiness INspector; FM = foundation model ; GNN = Graph Neural Network; KG = knowledge graph; LinkML = Linked data Modeling Language; LLM = Large Language Model; MCP = model context protocol; OSDR = NASA Open Science Data Repository; RadLab = OSDR's RadLab radiation data repository; RAG = retrieval augmented generation; SANS = Spaceflight Associated Neuro-Ocular Syndrome; scGPT = Single-Cell Generative Pre-trained Transformer; SPOKE = Scalable Precision Medicine Open Knowledge Engine; TabPFN = Tabular Prior-data Fitted Network.

Due to the challenges discussed above (data scarcity, heterogeneity, etc.) space life sciences data requires keen attention to AI-readiness at all points of the data lifecycle. This Perspective defines what AI-readiness demands (standardized metadata schemas, provenance tracking, quality metrics[32]) and suggests solutions for closing the gap between archived datasets and AI consumption. Using NASA's OSDR as a central example[2], we outline a path from existing repositories toward AI agent-accessible, analysis-ready infrastructure supporting space life science discoveries and the autonomous biological research deep space exploration will require.

## From FAIR to AI-Ready

Parallel international initiatives are establishing systematic data standardization and harmonization to enable data analysis across countries and data systems, including the International Standards for Space Omics Processing (ISSOP) consortium[33], the JAXA Integrated Biobank for Space Life Science[3], and the Space Omics Topical Team funded by ESA[34], each aligning with the foundational FAIR principles (findable, accessible, interoperable, and reusable[35]). Notably, an expansion of ISSOP or similar cross-organization coordination will be needed for AI-readiness itself, putting agencies, commercial operators, and academic partners on a shared footing for metadata schemas, provenance, quality scoring, and governance by providing an open forum for critical discussions among space biology experts and users.

While FAIR makes spaceflight data findable and reusable by human researchers, AI-ready data requires additional preparations[32]. Here, we treat AI-ready as an extension of FAIR. For example, the FAIR-R framework proposes expanding traditional data management to include AI-**R**eadiness (the ‘R’), emphasizing standards like provenance and structured labeling to ensure data is optimized for training[36]. For space biology, we present a functional hierarchy in which FAIR provides the foundation for AI-readiness, which in turn enables programmatic query and high-throughput training. The hierarchy culminates in space-readiness, which we define as data that AI can act on safely under spaceflight constraints (**Figure 1**). This hierarchy highlights the cost of skipping levels: without the AI-ready layer training is blocked; without the space-ready layer, models generalize poorly across unique spacecraft environments and small-*n* cohorts; and without the full stack, autonomous biological research in deep space stays out of reach.

The AI-ready tier is operationalized through frameworks such as AIDRIN (AI Data Readiness Inspector)[37], which adds the machine-consumable characteristics required for programmatic analysis. Broad AI-readiness initiatives are proliferating across the biomedical and governmental landscapes, from the NIH Bridge2AI framework[38], the BBSRC-funded network AIBIO-UK and genomic standardization via the Global Alliance for Genomics and Health[39], to cross-agency federal efforts such as the DOE-led Genesis Mission[40].

The Space-ready tier requires addressing five spaceflight-specific dimensions: small-*n* robustness, mission-context metadata, operational provenance, cross-mission harmonization, and governance. Examples of these constraints include restricted sample sizes due to mission manifest, biological readouts coupled tightly to operational conditions (e.g., $CO_2$, radiation), heterogeneity of methods across missions that requires harmonization and normalization, and restricted-tier biospecimen and cohort data that require governance-aware access. Implementation of this stack is underway through NASA's Transform to Open Science initiative[41].

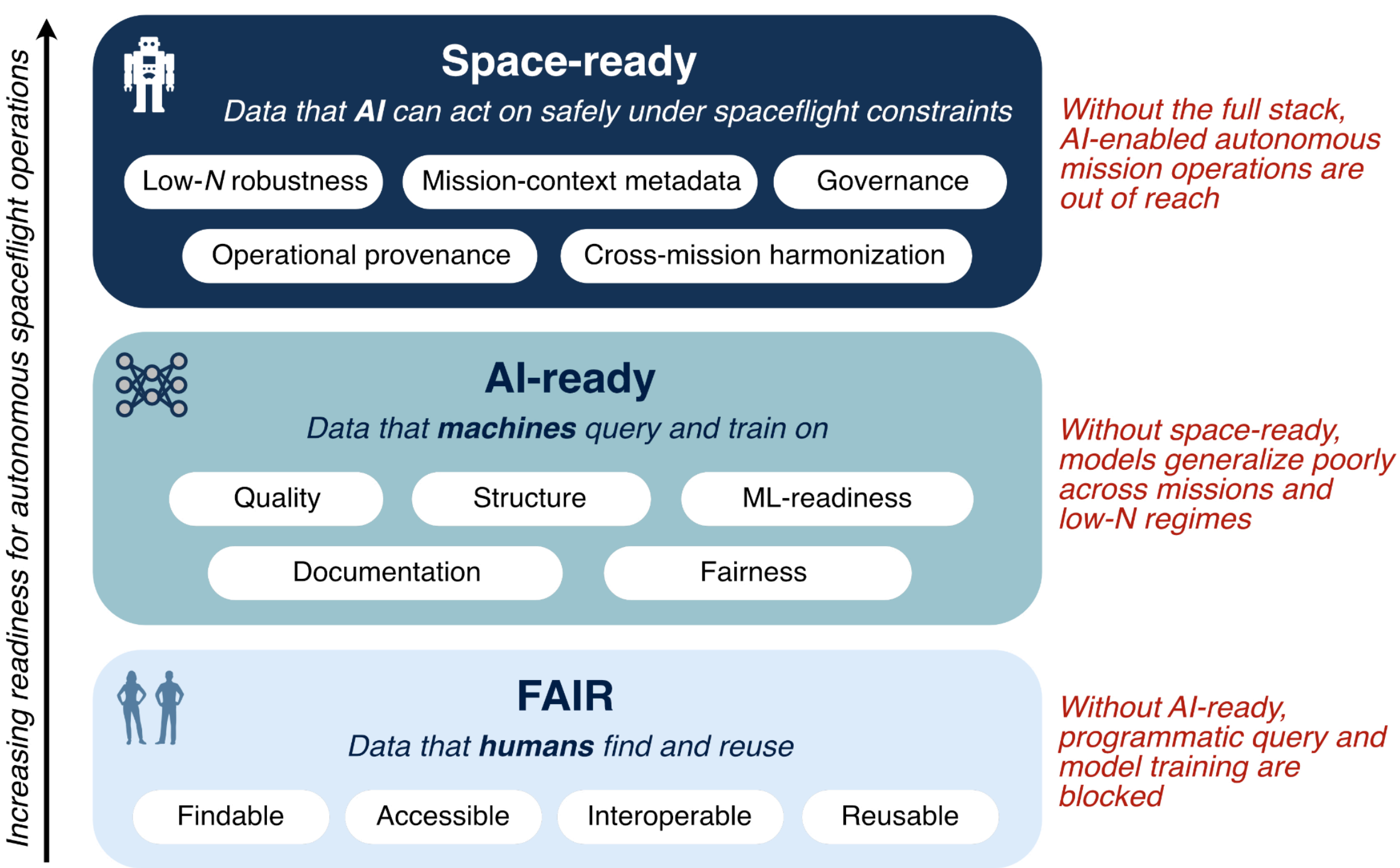

**Figure 1. Three-tier readiness approach for autonomous spaceflight operations**. Reliable AI for space life sciences and deep space exploration requires data that are FAIR, AI-ready, and space-ready, each tier contingent on its predecessor.

## AI-Readiness requirements

*Data characteristics across modalities*

AI-ready datasets require focus on specific quality markers: high label quality, harmonization (standardizing variables across different studies), and metadata richness to provide essential context for reuse. Further, datasets must maintain explicit provenance (i.e., a documented data's origin) and a principled handling of missingness to prevent model bias[42].

The choice of format for data storage is equally critical. For example, Apache Parquet, which stores data in columns rather than rows to allow for high compression and faster retrieval on disk, is often used by modern ML pipelines[43]. For active processing, Apache Arrow is the Parquet's in-memory counterpart, which enables ML pipelines to ingest and move data between tools without the need for time-consuming format conversion[44].

To accelerate the transition of repositories like the OSDR toward AI-readiness, we propose the adoption of a dedicated data layer based around the Parquet format. This would eliminate the "conversion tax" that currently plagues space biology due to the small-*n* datasets and high heterogeneity across hardware platforms and missions. For example, Rodent Research missions often return tissues from only tens of animals distributed across multiple investigators, while analog studies (e.g., bed rest, HERA, Antarctic) provide larger *n* at the cost of confounded variables like microgravity, radiation, and confinement.
This Parquet-based data layer would utilize massive Parquet tables partitioned by organisms ready to consume by any computing tools, including AI. For example, "All Mus musculus Transcription" could be collapsed into one single table using a Master Schema that maps metadata for alignment. Such an approach leverages OSDR's existing infrastructure to align disparate terms (e.g., mapping both "Space Flight" and "Flight" to the Space Life Sciences Ontology - SLSO_0000055), providing researchers with pre-aligned, machine-ready tables of expression matrices, telemetry, and metadata alongside legacy CSV/XLSX formats. This conversion is currently done manually, occupying hundreds of researcher hours and resulting in a high risk of error. Removing the conversion bottleneck would then allow researchers to leverage high-speed, in-memory processing via Apache Arrow for complex normalization and batch correction.
While this infrastructure provides the "tracks," the choice of normalization and harmonization remains with the investigator, and would be delivered extremely rapidly through the proposed data schema above. The field of space biology and omics is consistently changing and several approaches for such data harmonization have been explored[45]. For example, Casaletto et al.[30] applied harmonization to identify transcriptomic features associated with lipid density in spaceflown murine liver and age-dependent transcriptomic response in murine mammary tissue[46]. For modalities still rare in spaceflight (single-cell and spatially resolved 'omics), sample collection constraints, fixation protocols, and cross-study integration approaches all shape what AI workflows can support[47]. We identify three complementary paths beyond the single-study challenge: cross-study integration, transfer learning from terrestrial cohorts, and FM pre-training or fine-tuning, each with distinct data structuring requirements (**Table 1**).

*AI-ready metadata and curation*

OSDR uses the Investigation/Study/Assay (ISA) tabular format (ISA-Tab) for metadata[48] and community-vetted ontologies, ensuring consistent data characterization across the research community while capturing unambiguously investigation, study, and assay structure for all studies[2]. We recommend extending this format with git-style versioning and automated validation to support the scale of curation AI pipelines demand; migration to the ISA-JSON format could be considered to enhance programmatic access. Metadata standardization remains the principal bottleneck in scaling AI-ready infrastructure: hundreds of studies, thousands of assays, and diverse free-text fields require mapping to controlled vocabularies and ontologies. Recent advancements in LLMs and transformer architectures provide a path toward mitigating the intensive manual labor required to populate such granular metadata. For example, OntoGPT enables the extraction of structured terms from the unstructured text typically accompanying investigator submissions, aligning these elements with established ontologies[49]. Other tools like CurateGPT integrate retrieval-augmented generation with the LinkML schema framework[50] to support multi-step biocuration from schema design through evidence retrieval[19]. Both examples sit within the broader AI4Curation initiative, an open collection of LinkML- and ontology-based AI-assisted biocuration tools[51].
To streamline curation, we advocate for an LLM-assisted pipeline that uses ontology-focused tools like OntoGPT and CurateGPT, alongside general-purpose, schema-driven extractors like LangStruct, to extract structured metadata from raw text submissions. By integrating these with the LinkML schema framework, future repositories could automate the heavy lifting of data entry and schema validation. This would shift the curator's role from manual transcription to expert oversight and "edge-case" adjudication. Finally, AI Readiness frameworks such as AIDRIN[32] would provide a quantitative score for the resulting data quality and ML-readiness[32]. Together, this "closes the loop" between metadata curation and model performance, providing the rigorous audit trail now required by funders and journals.

*Provenance and workflow standards*

Provenance, the origin and processing history of each data transformation, is foundational to AI-readiness but weakly enforced by FAIR alone. While resources such as the W3C PROV-O ontology[52] provide the underlying schema for capturing lineage, licensing, and persistent identifiers, traditional data submissions often omit lineage documentation because it is manually cumbersome.
There are several steps in data provenance that must be well documented. Data curation is the first step in dataset publication, often performed by the data submitter or professional curators. The rise of autonomous AI agents in the curation process creates a new opportunity for rigor. While human curators often make subjective judgments, AI agents (e.g., AI4Curation tools) can document their internal reasoning chains, transforming curation from an opaque process into a transparent, machine-readable audit trail ecosystem[51].
Data processing often follows data curation, especially for large omics datasets, to ensure accessibility to scientists without strong computational biology skills. As such, data processing allows better reproducibility and interpretability of individual experiments. More generally, community standards have consolidated around data processing workflow-level FAIR principles[53]. However, to bridge the gap between documentation and institutional trust in the age of AI-enabled synthetic data generation, we also advocate for an authenticated provenance layer. By utilizing distributed ledger technology (blockchain) or cryptographic digital signatures, space agencies should provide a "digital seal" that certifies the authenticity and integrity of datasets as they move across international repositories. A current, more modest implementation is seen in NASA GeneLab's use of Nextflow[54] to ensure processing provenance travels natively with the data.
However, data integrity is only half the battle; we must also standardize how AI and ML model training and usage are reported. For this, we advocate that the space biology field adopt the DOME recommendations (Data, Optimization, Model, Evaluation)[55]. DOME provides a community-validated template for documenting the supervised ML process, ensuring that the evaluation and optimization steps are as transparent as the data provenance itself. This aligns directly with AIDRIN's ML-readiness pillar by providing the necessary "report card" for the model. At the policy layer, NASA's Open Science and Data Management Plan already couples funding to these requirements[56]. By introducing agent reasoning chains, agency-certified trust anchors, and DOME-standardized reporting, we transform metadata from mere description into verifiable evidence of a dataset's history and its subsequent utility in AI.

**The Access Layer**

AI-ready space biology data requires a multi-tiered access stack to ensure high-performance ingestion by downstream analytical tools. In this architecture, data holdings are synthesized through schema-driven curation into a robust KG and cataloged in a machine readable data registry, which is then interfaced with AI agents via MCP endpoints (**Figure 2**). The registry serves as the core metadata storage system at the stage of dataset curation, enabling the tracking of ETL lifecycles, data lineage information and AI-readiness metrics. This layered approach transitions data from static archives into dynamic, queryable resources. In the following section, we discuss key technical requirements for a tiered hierarchy, including KGs, MCP-mediated access, and the adaptation of FMs. We benchmark current OSDR capabilities against these future infrastructure targets. The same architecture can extend to additional space biology repositories including SOMA, LSDA/LSAH, HREDA, ibSLS, and TRISH EXPAND as repositories adopt interoperable metadata standards and AI-ready infrastructures.

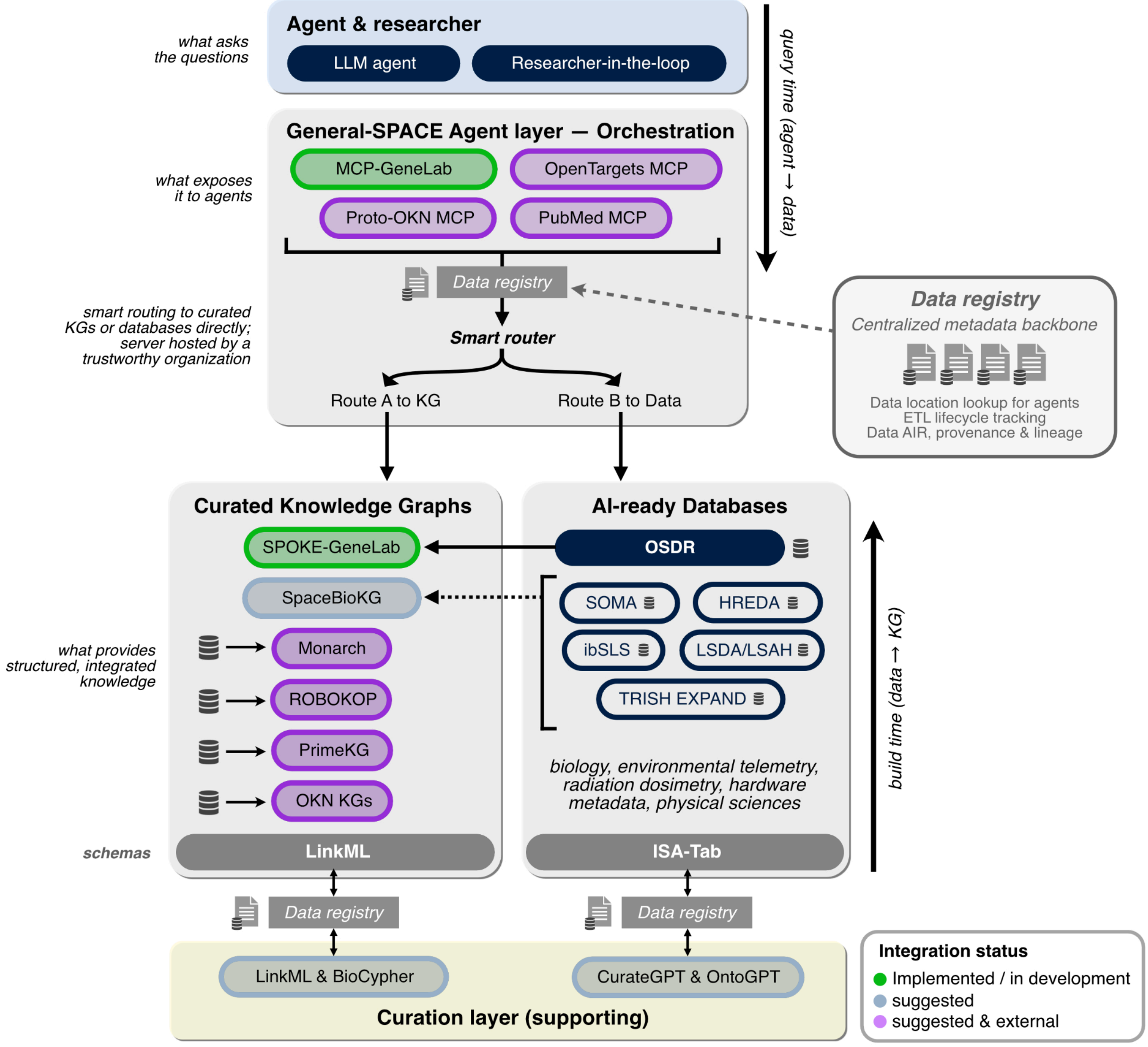

**Figure 2. Conceptual access-layer stack for transforming space biology repositories into agent-accessible, AI-ready knowledge ecosystems.** The architecture progresses from a supporting curation layer feeding into AI-ready databases, into curated KGs that integrate knowledge. At the orchestration layer, AI-ready data sources are cataloged into a machine readable data registry, and MCP servers provide intelligent routing depending on query context. The data registry also enables tracking ETL lifecycles and storing metadata and AI-readiness metrics. HREDA: Human and Robotic Exploration Data Archive; ibSLS: JAXA Integrated Biobank for Space Life Science; KG: knowledge graph; LSAH: Lifetime Surveillance of Astronaut Health; LSDA: NASA Life Sciences Data Archive; MCP: Model Context Protocol; Proto-OKN: NSF Prototype Open Knowledge Network; AIR: AI-Readiness.

*Knowledge Graphs*

KGs represent entities such as genes, proteins, phenotypes, missions, hardware systems, and environmental conditions (nodes) connected by relationships (edges), enabling AI systems to reason across heterogeneous biomedical and spaceflight data through semantically linked, multi-relational inference. Unlike traditional databases that primarily store observational evidence, KGs integrate relationships across domains to support hypothesis generation, contextual reasoning, and cross-dataset discovery.

Biomedical KGs including the Monarch Initiative[57], SPOKE[58], CROssBAR[59], ROBOKOP[60,61] and PrimeKG[21] demonstrate the value of graph-based integration for translational biomedicine by connecting genes, diseases, phenotypes, drugs, pathways, and clinical concepts across multiple knowledge sources. ROBOKOP, for example, focuses on explainable biomedical reasoning through graph traversal and question answering, while PrimeKG emphasizes integrated disease–gene–drug relationships for precision medicine applications. The approach also extends naturally to spaceflight biology: Nelson and colleagues embedded GeneLab transcriptomic data from six spaceflown-mouse studies into SPOKE, recovering disease signs and symptoms that differential expression alone could not surface[20]. The SPOKE-GeneLab KG has been expanded to include OSDR RNA-seq, methylation sequencing, and amplicon sequencing data, alongside metadata nodes that can be used as connectors to external KGs, including NSF's Proto-OKN, enabling the integration of spaceflight experimental results with ground-based biomedical resources. We propose extending this paradigm further across OSDR's broader multi-omics and environmental catalog, allowing AI agents to reason across missions, hardware platforms, environmental stressors, and radiation environments while preserving provenance links between every node, edge, dataset, and supporting publication.

To support this integration, standardized metadata and schema frameworks are required at multiple levels of the ecosystem. AI-ready databases such as OSDR rely on structured metadata standards including ISA-Tab to harmonize omics, environmental telemetry, radiation dosimetry, and mission-associated experimental data. In parallel, KG-oriented schema and graph-construction frameworks such as LinkML[50] and BioCypher[62] facilitate the transformation of heterogeneous biomedical datasets into interoperable graph representations aligned with FAIR principles (**Figure 2**).

*Model Context Protocol servers*

MCP provides a contextual orchestration layer that dynamically selects and composes the most relevant resources for a given query, including curated KGs, AI-ready databases, and external biomedical services. BioContextAI aggregates biomedical MCP servers, providing the connective infrastructure through which agents can discover, route, and integrate biomedical analyses across heterogeneous resources[63].

Building on the SPOKE-GeneLab integration described above, the OSDR team is developing the open-source MCP-GeneLab server (https://github.com/sbl-sdsc/mcp-genelab) to enable natural-language queries against the SPOKE-GeneLab KG (https://github.com/BaranziniLab/spoke_genelab), exposing space-relevant transcriptomics, epigenomics, and metagenomics data, and associated metadata through a chatbot client and AI agents. External MCP servers including the NSF MCP-Proto-OKN (https://github.com/sbl-sdsc/mcp-proto-okn), PubMed MCP, and Open Targets MCP (https://mcp.platform.opentargets.org/) further extend the ecosystem by enabling federated reasoning across spaceflight and terrestrial biomedical knowledge sources through natural-language queries alone.

Rather than exposing individual repositories through isolated MCP endpoints, we envision a generalized agentic orchestration layer ('General-SPACE Agent') that provides context-aware routing between AI-ready databases, curated KGs, and external biomedical resources (Figure 2). In this framework, an AI agent acts as a trusted mediation layer, selecting and invoking the appropriate MCP-connected tools based on query context. Depending on the nature of the query, the agent may invoke semantic reasoning over KGs, direct retrieval from observational datasets, or federated integration with external biomedical services through tools such as PubMed MCP, Open Targets MCP, or Proto-OKN. To facilitate agent orchestration and enable agentic findability for all data sources, we propose a machine readable data registry that contains descriptive metadata and routing information for each AI-ready database that is exposed through an API and callable via an agent tool or skill.

A critical aspect of this architecture is governance. We propose that MCP services be hosted or validated by trusted neutral organizations capable of enforcing provenance tracking, tool validation, access control, auditability, and safety constraints for agentic workflows. This becomes particularly important in biomedical and spaceflight applications where AI agents may autonomously compose analyses across heterogeneous resources with varying evidence quality and operational sensitivity. Under this model, General-SPACE Agent provides a scalable mechanism for integrating biological measurements, radiation dosimetry, environmental telemetry, mission metadata, and external biomedical knowledge into a unified agent-accessible ecosystem while preserving transparency and scientific traceability. The same orchestration pattern can generalize across additional space biology repositories as their AI-readiness matures.

*LLM retrieval patterns*

For unstructured and semi-structured text, such as OSDR study descriptions, methods sections, or the broader biomedical literature, RAG can identify passages that semantically match a query rather than relying solely on traditional database searches over metadata fields or tabulated parameters. In graph-based retrieval settings, however, conventional vector-based RAG alone becomes insufficient because it flattens the relational structure, ontology hierarchy, and provenance relationships that make graph-based representations scientifically valuable. Alternative retrieval strategies are therefore needed. One approach involves agentic tool orchestration over MCP-connected resources, where an agent dynamically selects the appropriate tools, databases, or KGs depending on query context. For example, an agent with access to MCP-exposed graph query interfaces could translate a natural-language question into a formal graph query language such as Cypher or SPARQL, allowing direct interrogation of graph structure and relationships[64] (Example 1).

**Example 1: Converting a natural-language question into a graph query**

User asks: “Which genes associated with mitochondrial dysfunction were upregulated in mice exposed to space radiation?”

The system translates this into a Cypher query (Neo4j graph language):

</> cypher

```
MATCH (g:Gene)-[:ASSOCIATED_WITH]->(p:Pathway {name:"Mitochondrial Dysfunction"})
MATCH (g)-[:UPREGULATED_IN]->(e:Experiment {name:"Space Radiation"})
RETURN g
```

*Note: The above example is simplified for illustration.*

Another strategy is KG-RAG, which retrieves biologically relevant subgraphs from curated KGs[18] by semantically mapping user queries onto entities and relationships within an existing graph structure. Unlike conventional RAG, which retrieves independent text passages, KG-RAG preserves explicit biological relationships between genes, pathways, phenotypes, missions, environmental exposures, and experimental conditions, enabling more interpretable and provenance-aware inference.

This distinction is important because recent popular GraphRAG approaches, such as Microsoft's GraphRAG framework[65], generate graph abstractions dynamically from vectorized documents using LLM-based entity and relationship extraction. While effective for summarization and exploratory retrieval over unstructured text, these induced graphs differ fundamentally from our proposed usage of curated biomedical KGs such as SPOKE, Monarch, ROBOKOP, or PrimeKG. In this latter case, graph topology, ontology alignment, and evidence provenance are explicitly curated and biologically grounded. Figure 2 emphasizes the importance of hybrid architectures combining conventional RAG for literature retrieval with curated KG reasoning for mechanistic and translational inference.

*Foundation Models*

In the previous section we discussed new retrieval strategies. Here, we emphasize how these approaches become most powerful when paired with FMs capable of reasoning across heterogeneous retrieved contexts. FMs address two structural realities of space biology: small per-study sample sizes (*n*), which limit training from scratch, and strong per-mission heterogeneity that challenges single-task models[8,9]. Transfer learning using FMs trained on large-scale biological datasets enables space biology applications to leverage representations that have already captured many core biological processes while requiring far fewer spaceflight datasets to achieve useful predictive performance. Recently, the biomedical AI community has cautioned against the rapid proliferation of fragmented and partially redundant biomedical FMs lacking standardized benchmarking, transparency, and clear deployment pathways[66]. Accordingly, rather than developing isolated “space-specific” models by default, space biology should initially prioritize adapting, benchmarking, and fine-tuning existing validated FMs before attempting to pre-train new models from scratch.

For example, tabular FMs such as TabPFN are relevant because they were designed for small-to-medium tabular datasets and can make strong zero-shot predictions in a single forward pass, outperforming tuned gradient-boosted decision trees in benchmark tasks[16]. For OSDR’s gene expression count matrices, environmental telemetry, and structured spaceflight data, this makes tabular FMs a promising first AI-assisted analysis path for the small-*n* regime. Similarly, terrestrial omics FMs can be fine-tuned for spaceflight-specific tasks. Geneformer used transfer learning from approximately 30 million single-cell transcriptomes to make context-specific predictions in network biology[14], while scGPT was trained on more than 33 million cells to learn generalizable representations of genes and cells for single-cell and multi-omics analysis[15]. These training datasets are at least one order of magnitude larger than OSDR’s current single-cell catalog, making pre-training a space-biology FM from scratch unrealistic at present. Fine-tuning requires substantially less data and is increasingly supported by agentic biomedical platforms such as Biomni, which integrates LLM reasoning, retrieval-augmented planning, and code-based execution to automate biomedical research workflows[67].

Three requirements remain critical: First, batch-effect-aware harmonization across missions and platforms; second, integration of mission-context metadata into training pipelines; and third, privacy-aware training methods that respect astronaut data-access restrictions, since small cohort sizes make de-identification alone insufficient against re-identification[12]. Federated learning, discussed in the next section, addresses the latter challenge.

*From Research Data to Operational Use*

The access-layer infrastructure for research described above is equally relevant to operational crew health, where the same KGs, MCP-connected tools, and federated-learning approaches would support biomonitoring, clinical decision support, and life-support diagnostics under communication latency. The data classes, regulatory pathways for medical software in flight, and governance regimes that operational use requires sit outside the scope of this Perspective; early operational-facing efforts include HERMES[68], the proposed Human Cell Space Atlas[69], and cross-archive frailty signature work[70]. Translating this infrastructure into operational crew health will require a separate, stakeholder-driven analysis, one that convenes flight surgeons, NASA's Office of the Chief Health and Medical Officer (OCHMO) and Crew Health and Safety leadership, Environmental Control and Life Support Systems (ECLSS) and life-support engineers, space medicine researchers, commercial-provider medical officers, and international-partner operational groups to define the data classes, regulatory pathways, and governance models appropriate to that setting.

### Federated learning and privacy-preserving analysis

A central obstacle to using ML in biomedical research in general and space biology specifically is the fragmented nature of the relevant datasets, spread across organizations, institutions, and experimental platforms. Many contain sensitive information or sit behind access restrictions that prevent centralized aggregation[12]. Federated learning offers a path forward: rather than transferring privacy-sensitive data to a model, the model is transferred to each data silo. Each site trains a local model on its own data and shares only the resulting parameter updates (i.e., gradients or weight deltas) with a central aggregator, which merges them into a refined global model (e.g., via weighted averaging or FedAvg) and broadcasts it back to all participants. This round-trip repeats until the global model converges across all sites, and raw data never need to leave their system of record[71].

The Federated Learning using in-space Data (FLUID) framework is, to our knowledge, the first federated learning framework reported to run aboard a crewed spacecraft on biological data [23,72]. Built on the OpenFL implementation[82] and validated on HPE's Spaceborne Computer-2 aboard the International Space Station (ISS), with the aggregator operating Earth-side, FLUID trained a machine learning causal-inference ensemble on cardiac radiation response data sourced from OSDR, reaching accuracy comparable to centralized training under realistic communication latency and loss-of-signal events. A dedicated shim layer managed asynchronous parameter exchange across intermittent communication windows, a design constraint that any future spaceflight federation will need to replicate.

We propose extending this approach across the spaceflight biological data ecosystem (**Figure 3**) through a three-zone architecture. Zone 1 would encompass the data silos themselves: i.e. NASA and commercial spaceflight cohorts, human analog studies, tissue-on-a-chip platforms, and large terrestrial biobanks. Each silo would run a local ontology mapper, an AI Readiness quality scorer, and an FL collaborator. Critically, each silo would first map its native schema to a shared ontological standard via a local KG (e.g., LinkML, SPOKE), ensuring that local models across disparate platforms train on comparable feature representations before any parameter exchange occurs. This upstream harmonization step is a prerequisite for meaningful aggregation: without feature alignment, model performance degrades. AIDRIN[32] could be used to score each participating dataset against quality, fairness, and ML-readiness criteria before a site enters the federation, and again after global model aggregation, providing a continuous readiness gate.

Zone 2 is the neutral orchestration entity, proposed to be hosted by the same trusted independent body or organization hosting the General-SPACE Agent introduced in Figure 2. This body would host the Earth-side aggregator providing the governance layer for the entire system. This entity would receive differentially private gradient updates from all collaborators, perform secure aggregation, and broadcast the updated global model back to every site for the next training round. It would also maintain a verified MCP tool registry, ensuring that only audited, signed tools can be invoked by downstream agents, and an immutable provenance ledger recording which datasets were used to train each model. Collocating federation governance with agent governance in a single neutral entity is deliberate: it creates one auditable trust boundary rather than two.

Zone 3 is the query and access layer. The converged global model's outputs and provenance metadata would pass through the same curation pipeline described above (i.e., LinkML, BioCypher, OntoGPT) populating the federated KG and AI-ready databases that AI agents query via a smart router using MCP-connected tools (**Figure 2**). KGs would thus serve a dual role in this architecture: upstream, as a per-silo harmonization layer that makes federation possible; and downstream, as the structured semantic surface through which federation-level insights become accessible without exposing any underlying data.

FL is not without limitations. It is particularly vulnerable to datasets that are not independent and identically distributed (IID), resulting in lower accuracy, delayed convergence, and unfairness. In our proposed solution, the upstream KG harmonization step mitigates but does not eliminate these risks, and extreme residual heterogeneity can still degrade global model quality below what a single high-quality site would achieve. FL also introduces its own attack surface: gradient inversion and membership-inference attacks can partially reconstruct training data from parameter updates, requiring secure aggregation protocols and calibrated differential privacy noise as complementary safeguards. Finally, FL preserves data sovereignty at the compute layer but does not substitute for the legal and institutional infrastructure that must surround it: data-use agreements, protocol review, and cross-jurisdiction governance frameworks remain necessary. FLUID demonstrates that the compute layer is feasible in the spaceflight environment; the governance infrastructure for a multi-agency and commercial federation remains to be established.

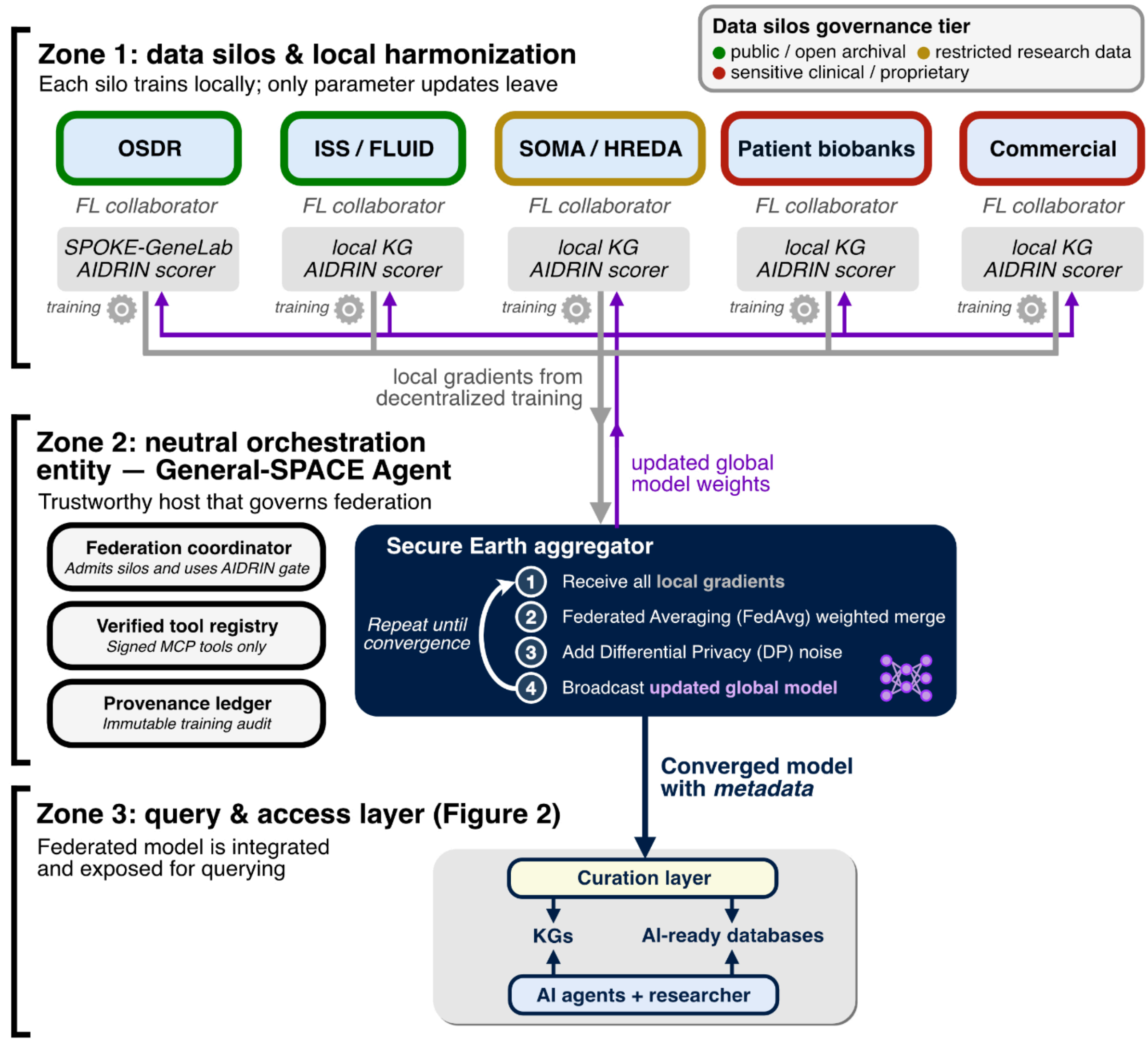

**Figure 3. Proposed three-zone federated learning architecture for the space biological data ecosystem.** A federated learning (FL) architecture enables AI models to train across diverse, distributed data sources without centralizing sensitive data. Each peripheral node in Zone 1 represents a data source category contributing to the federation; local parameter updates (gradients) travel upward to the Earth aggregator in Zone 2, while raw data remain within their origin silo. The aggregator iteratively merges updates until the global model converges across all sites. Zone 2 is governed by the General-SPACE Agent, an agentic orchestration component that manages federation eligibility via an AI Readiness quality scoring framework such as AIDRIN, enforces differential privacy during aggregation, maintains an immutable provenance ledger, and restricts agent tool calls to a registry of audited, signed MCP tools. The converged global model feeds Zone 3, where a curation layer maps outputs to a standardized ontological schema, populating curated KGs and AI-ready databases accessible to LLM agents and researchers via a smart query router as introduced in Fig. 2.

**Infrastructure and interoperability**

Space agencies and commercial operators currently maintain parallel biological data infrastructures. NASA's Life Sciences Data Archive at Johnson Space Center catalogs astronaut research data with a public request pathway, and Bloomfield et al.[73] proposed an international expansion of its longitudinal scope. JAXA operates the publicly accessible ibSLS repository and biobank[3] alongside their orbital experimental infrastructure[74]. ESA's Human and Robotic Exploration Data Archive catalogs ISS and microgravity ground experiments. In parallel, commercial providers (e.g., Axiom Space, SpaceX, Blue Origin) maintain internal infrastructure with limited external visibility. Two recent initiatives have begun surfacing commercial and multi-mission astronaut biological data for outside researchers: the Space Omics and Medical Atlas (SOMA), spanning the NASA Twins Study, JAXA CFE, SpaceX Inspiration4, Axiom, and Polaris missions, with matched samples through the Cornell Aerospace Medicine Biobank[5]; and TRISH's Enhancing eXploration Platforms and ANalog Definition (EXPAND) Database and Biorepository, opened in March 2026.

AI-readiness varies considerably across these holdings, including data format, metadata completeness, access pathway, and ML-readiness criteria. This makes cross-infrastructure comparability difficult to assess and harder still to guarantee. Realizing the federated architecture proposed here requires a neutral international coordinating body, independent of any single space agency or commercial operator, with the mandate to audit datasets against common standards, certify tool interoperability, and enforce the governance infrastructure that no individual stakeholder has the authority or incentive to impose unilaterally. The formation of this body could be modeled on the International Standards for Space Omics Processing (ISSOP) consortium[33] but with a specific mandate focused on AI support, including enforcing common AI-readiness standards, tool certification, and cross-jurisdictional governance across all stakeholders.

Three interlocking standards are necessary to build integrated AI-ready infrastructure across these holdings. Common metadata schemas, established by ISA-Tab and extended through LinkML as discussed above, provide the shared vocabulary on which the other layers depend. Federated discovery protocols such as the Global Alliance for Genomics and Health (GA4GH) Beacon v2 demonstrate that private data can be searchable without being accessible, and this is deployable at an international scale[75]. This provides a governance template for the broader federation we are proposing here. AI-readiness scoring through AIDRIN provides the third standard, ensuring that any dataset entering a cross-organization federation meets consistent criteria for downstream model use[32]. On top of these, ML-specific dataset descriptors (Croissant, Datasheets for Datasets, Data Cards) expose dataset-level information tuned to ML consumption[76,77]. We recommend OSDR adopt Croissant export for its AI-ready datasets to align with downstream ML tooling and the NeurIPS/ICML dataset-track expectations, increasing the visibility of spaceflight data in the AI community– complementary model-artifact metadata standards (AIO, MLLO) are emerging in parallel.

A recent ESA-NASA collaboration ingested ESA's *Adineta vaga* spaceflight transcriptomics study (OSD-824) into OSDR's Nextflow RNA-seq pipeline with full ISA-Tab metadata[78,79], demonstrating a concrete step toward globally integrated AI-ready infrastructure. Together with the access layer and federated learning described above, this infrastructure would let AI agents query, train on, and reason across the entirety of space biology data without compromising the sovereign constraints that currently fragment the field.

Nevertheless, the value of this infrastructure depends on who can use it. **Supplementary Table 1** maps eight representative user communities (AI/ML developers, space life sciences researchers, flight surgeons, ECLSS systems engineers, mission planners, commercial spaceflight operators, students and trainees, and citizen scientists) onto the access-layer stack, identifying the primary access mode, key resources, and consequences of incomplete AI-readiness for each.

**Outlook and opportunities**

Space biology AI-readiness metrics need to mature from qualitative checklists into quantitative, dataset-level scores that funders, journals, and downstream model developers can act on. Several frameworks are converging on this problem, such as AIDRIN scores[32], SciHorizon benchmarks[80] and the Bridge2AI Standards Working Group[38]. Selecting, reconciling, and enforcing a common scoring standard across the ecosystem is precisely the kind of mandate that cannot be left to individual archives or agencies, and should be the responsibility of the neutral international coordinating body proposed above.

The practical payoff of that standardization is already visible: NASA's release of AI benchmark training datasets on the AWS Registry of Open Data shows that richly labeled, unambiguous data transforms archived experiments into resources the research community can immediately act on[2]. The infrastructure proposed in this perspective extends that logic to the entire space biology ecosystem, i.e., a unified KG schema and standardized access layer would make every AI-ready dataset a potential benchmark by default, without requiring per-dataset curation effort.

The relationship can then become bidirectional. Researchers who train on these benchmarks contribute pre-trained model artifacts back to the ecosystem, hosted alongside the source data so that model performance metrics and data quality scores can be evaluated together. These models become the foundation for transfer learning and fine-tuning on new missions and datasets, compounding value over time rather than restarting from scratch with each new experiment. There are several ML distribution platforms, such as Hugging Face Hub and the AWS Registry of Open Data, ensuring accessibility and version control[81]. In such distributions, terrestrial biomedical FMs such as Geneformer, scGPT, and TabPFN are already shared, ensuring the space biology community builds on rather than alongside the broader AI ecosystem.

Agentic AI systems extend the access layer by navigating heterogeneous data structures, orchestrating tool use across APIs, and composing multi-step analyses without manual workflow construction[17]. Their value depends on quality metrics for unstructured data, explicit provenance tracking, and empirical correction for the small sample sizes and mission-specific biases of spaceflight experiments. Governance remains a limiting factor rather than a facilitator: existing ethical and policy frameworks for astronaut genomic and health data were not designed with federated AI in mind, and overly conservative interpretations of policy have been driven by risk aversion rather than evidence of harm. As a result, access to data whose scientific and operational value is substantial is delayed[82]. Commercial spaceflight compounds this with a separate regulatory vacuum, where intellectual-property terms, commercial-crew agreements, and the absence of consensus ethics-review frameworks further limit what can train open models[83]. The field needs frameworks proportionate to actual risk, guided by people who understand both the data and the technology. Closing that gap requires coordinated action: funding mechanisms that treat AI-readiness as a deliverable alongside scientific results, cross-agency and public-private agreements that convert data sovereignty constraints into federated solutions rather than barriers, and a neutral international coordinating body with the mandate to enforce common standards across the ecosystem. Without governance at that level, the infrastructure described in this perspective remains a set of prototypes rather than a deployed system.

The Artemis program and commercial spaceflight will generate more biological and environmental data than any prior era of human space exploration. Without AI-ready infrastructure, that data will primarily be archived and may be accessible, but not used at its full potential. In this perspective, we showed how the transition from FAIR to AI-ready is not a technical option but a mission-enabling requirement. We identified the key emerging components: i.e., KGs, agent-accessible query layers, federated learning demonstrated aboard the ISS, end-to-end pipelines, and LLM-assisted curation. The task now is integration at scale across NASA, international partners, and commercial operators, on a timeline that matches the cadence of deep space exploration. The biological data that deep space exploration will depend on already sits in agency and commercial archives, where it continues to accumulate. The agencies, archives, and commercial operators that steward it now face a decision: whether to make it AI-ready.

**Acknowledgements**
The authors thank the members of the OSDR AI/ML Analysis Working Group for valuable input, and discussions and for their constructive comments. We are also grateful to Mona Nasser, Frank Soboczenski, Saswati Das, and Nathaniel Szewczyk for insightful discussions during the early stages of this project. S.V.C., A.B. and C.E.M are supported by the Trivedi Institute for Space and Global Biomedicine at the University of Pittsburgh. C.E.M. thanks the WorldQuant Foundation, the Katherine And John Bleckman Foundation, and NASA (80NSSC22K0254, 80NSSC23K0832, 80NSSC24K0728, 80NSSC25K7731), the National Institutes of Health (R01ES032638, U54AG089334, P01CA272295, R01CA266279). S.E.B. and P.W.R. were in part supported by the National Science Foundation Award #2333819: 'Proto-OKN Theme 1: Connecting Biomedical information on Earth and in Space via the SPOKE knowledge graph'. S.G. is supported by the Swedish National Space Agency (Rymdstyrelsen; 2025-00267, 2024-00179) and the Swedish Research Council (Vetenskapsrådet, 2024-05289) in relation to this work. S.G.B. is funded by a Wellcome Sanger Institute PhD Studentship. The Wellcome Sanger Institute is supported by core funding from the Wellcome Trust (grant 220540/Z/20/A). A.M.S., J.A.C., and R.T.S. are part of the 'AI for Life in Space' (AI4LS) group at NASA Ames, and are grateful for its support as well as that of the NASA Science Mission Directorate's 'Foundation Model' effort.

**Author contributions**
P.M. conceived the idea for this study and supervised the project. S.V.C. and P.M. wrote and iteratively revised the manuscript, with contributions from other authors. R.T.S. and S.V.C. led a substantial final revision addressing co-author feedback, and edited the recommendations for clarity and consistency. L.M.S. and S.V.C. prepared the submission-ready manuscript with contributions from A.B., R.T.S., J.A.C., B.M.E., S.G.B., S.G., C.E.M., P.W.R., S.E.B., and P.M. Figures and tables were designed and produced by R.T.S., S.G.B., S.V.C., and P.M. A.M.S. contributed intellectual content on KGs, MCPs, and provided critical feedback on LLM retrieval mechanisms for unstructured vs. structured data sources. All authors contributed to the writing, revised the manuscript, and approved the final version.

**Competing interests**
The authors declare no competing interests.

**Use of artificial intelligence**
During the preparation of this manuscript, the authors used large language model (LLM)-based tools — including Claude (Anthropic), Gemini (Google), and ChatGPT (OpenAI) — to support language editing, improve clarity of expression, and assist in articulating complex technical concepts. These tools were not used to generate original text, ideas, figures, or scientific content; all intellectual contributions, interpretations, and conclusions are solely those of the authors. All AI-assisted passages were reviewed, revised, and approved by the authors to ensure accuracy and fidelity to the intended meaning. This disclosure is made in accordance with the 2024 recommendations of the International Committee of Medical Journal Editors (ICMJE; www.icmje.org) and the position statement of the Committee on Publication Ethics (COPE; publicationethics.org/cope-position-statements/ai-author) on the transparent reporting of AI assistance in scholarly writing.

**Supplementary Table 1.** Use-case scenarios for an AI-ready space biology data ecosystem. Eight representative personas each engage a different region of the access-layer stack (Figure 2) and depend on specific combinations of AI-ready resources named in this Perspective. The rightmost column identifies what breaks for each persona when the full stack is not in place, matching the failure-mode framing of Figure 1.

| Example User | Representative use case | Primary access mode | Key resources | Result of lack of AI-readiness |
|---|---|---|---|---|
| **AI/ML developer or data scientist** | Training FMs and classical predictors on spaceflight multi-omics; developing domain-specific models for crew health outcomes; benchmarking methods on reproducible datasets. | Programmatic APIs and MCP endpoints | OSDR bioData API, SPOKE-GeneLab KG, MCP-GeneLab, AWS BPS benchmarks, AIDRIN scoring, Croissant metadata | Fragmented, unharmonized data defeat model training; small *n* blocks fine-tuning without cross-mission pooling; no standardized benchmarks for method comparison. |
| **Space life sciences researcher** | Cross-mission hypothesis testing on tissue-specific spaceflight responses; integrating omics with radiation and environmental context; comparing rodent and human signatures across missions. | Metadata schemas and KGs | OSDR, SOMA, ISA-Tab, LinkML, SPOKE-GeneLab, Monarch Initiative, PrimeKG | Metadata gaps and batch effects block cross-study pooling; missing controlled vocabularies defeat semantic queries; cross-organism comparison requires manual harmonization. |
| **Flight surgeon or aerospace medicine clinician** | Contextualizing crew symptoms and performance deviations against mission history, prior cohorts, terrestrial precedent, and occupational surveillance records; decision support during deep-space missions with communication latency; countermeasure selection. | Agent interfaces with governance guardrails | HERMES, TRISH EXPAND, LSAH (restricted), SOMA, Monarch MCP, PubMed MCP | Decision-support outputs cannot be grounded in provenanced evidence; regulatory pathways block deployment; autonomous triage impossible without auditable reasoning chains. |
| **Systems engineer (ECLSS and life support)** | Detecting hardware anomalies before crew health impact; correlating environmental telemetry with biological response signals; cabin-atmosphere and radiation monitoring. | Telemetry APIs and MCP servers | EDA, RadLab, MCP-EDA (planned), MCP-RadLab (planned), OSDR biological co-measurements | Operational telemetry stays siloed from biology; anomaly detection lacks biological context; radiation-biology co-analysis requires bespoke integration per mission. |

| Example User | Representative use case | Primary access mode | Key resources | Result of lack of AI-readiness |
|---|---|---|---|---|
| **Mission planner or program manager** | Risk stratification and countermeasure prioritization for lunar and Mars missions; crew readiness, performance, and mission-objective surveillance across mission phases; quantifying evidence gaps to target future research investment; AI-readiness assessment across holdings. | Aggregated readiness dashboards | AIDRIN readiness scoring, OSDR, SOMA, federated-learning aggregated insights, HERMES | Risk assessment falls back on anecdote; evidence-gap quantification impossible without machine-readable metadata across the ecosystem; countermeasure prioritization lacks statistical grounding. |
| **Commercial spaceflight operator or provider** | Pooling de-identified crew data across providers for cohort-scale inference while preserving commercial confidentiality; meeting emerging regulatory expectations for safety data sharing. | Federated learning with privacy guarantees | SOMA, FLUID federated-learning infrastructure, TRISH EXPAND, AIDRIN pre-federation scoring, GA4GH Beacon v2 | Competitive and regulatory barriers prevent cross-operator learning; privacy-preserving analysis tooling not available; no standard for pre-federation quality assurance. |
| **Student, trainee, or early-career researcher** | Learning multi-omics and ML methods on reproducible spaceflight benchmarks; thesis research on publicly available datasets; onboarding to the space biology community. | Benchmark datasets and tutorials | AWS BPS microscopy and RNA-seq benchmarks, OSDR tutorials, SOMA public tier, open-source MCP servers, GL4U training | Training materials fragmented across archives; reproducibility barriers block thesis-scale analyses; no low-friction on-ramp to cross-mission research. |
| **Citizen scientist or open-science contributor** | Participating in hackathons, community challenges, and reproducible analyses on publicly available spaceflight data; contributing annotations and method variants back to the ecosystem. | Public portals and open data registries | OSDR public access, AWS Registry of Open Data BPS datasets, SOMA public tier, AWG community forums | Entry barrier too high for non-institutional contributors; documented APIs and low-friction access paths absent; community engagement stays shallow. |